 \documentstyle[11pt,aaspp4]{article}

\begin{document}

\title{The N/Si Abundance Ratio in Fifteen Damped Lyman-alpha Galaxies:
Implications for the Origin of Nitrogen$^1$}
\author{Limin Lu$^{2}$, Wallace L. W. Sargent}
\affil{California Institute of Technology, 105-24, Pasadena, CA 91125}
\author{and Thomas A. Barlow}
\affil{Center for Astrophysics and Space Sciences, University of California,
San Diego, C-0424, La Jolla, CA 92093}
\altaffiltext{1}{Based partially on observations obtained at the W. M. Keck
observatory, which is jointly operated by the California Institute of
Technology and the University of California.}
\altaffiltext{2}{Hubble Fellow}

\begin{abstract}

Galactic chemical evolution model calculations indicate that
there should be considerable
scatter in the observed N/O ratios at a fixed metallicity (O/H) for galaxies
with very low metallicities due to the delayed release of primary
N from intermediate mass stars relative to that of O from short-lived
massive stars. Moreover, the scatter should increase progressively
toward decreasing metallicity. Such effects have not been
convincingly demonstrated by observations of H II regions in nearby
metal-poor galaxies, raising doubts about the time-delay model of
primary N production.  Pettini et al and Lipman et al realized 
the utility of high-redshift damped Ly$\alpha$ galaxies for gaining
further insights into the origin of N and discussed abundances in
three damped Ly$\alpha$ galaxies. Since abundance measurements
for O are generally unavailable for damped Ly$\alpha$ galaxies, 
they used N/Si or N/S in place of N/O under the reasonable assumption 
that the abundance ratios O/Si and O/S are
the same as solar in damped Ly$\alpha$ galaxies.
We discuss observations of heavy element abundances
in 15 high-redshift ($z>2$) damped Ly$\alpha$ galaxies, 
many of which have metallicities comparable to or lower than the
lowest metallicity galaxy known locally (I Zw 18). 
We find that the N/Si ratios in damped Ly$\alpha$ galaxies
exhibit a very large scatter ($\sim1$ dex)
at [Si/H]$\sim -2$ and there is some indication that the scatter 
increases toward decreasing metallicity. 
Considerations of various sources of uncertainties
suggest that they are not likely the main causes of the large scatter.
These results thus provide strong support for
the time-delay model of primary N production in intermediate mass stars
{\it if, indeed, O/Si$\simeq$solar in damped Ly$\alpha$ galaxies}.
   
\end{abstract}


\section{INTRODUCTION}

    The element Nitrogen is undoubtedly synthesized from Carbon and Oxygen
through the CNO 
cycle in stellar interiors. However, the details of the nucleosynthesis
-- the mass range of the stars and the  
stage of evolution -- are uncertain.
According to current theory, ``primary'' nitrogen production 
occurs in the Asymptotic Giant Branch phase of the
evolution of intermediate mass stars (3-8$M_{\odot}$), 
when thermal pulses bring up C-rich material from the He-burning 
shell into the hydrogen-burning shell (Renzini \& Voli 1981). 
The N subsequently synthesized in the hydrogen-burning
shell is later dispersed into the interstellar medium through stellar winds.
The N produced in this manner is considered ``primary'' because 
its production does not depend on the initial heavy element content of the 
stars in order to provide the seed C nuclei.  The synthesis of N can also 
occur through the CNO cycle in main sequence stars of any mass which
contain an initial supply of heavy elements. 
The amount of N produced in this latter manner depends on
the initial metallicity of the stars; hence this mechanism is 
described as  ``secondary''.  Galactic chemical evolution models predict
that the N/O abundance ratio should be independent of the overall heavy 
element abundance (N/O independent of O/H) for 
primary N production. On the other hand, 
it is expected that N/O $\propto$ O/H 
for secondary production (Pagel \& Patchett 1975; Edmunds 1990).  
For stellar systems with normal initial mass functions, there should 
necessarily be a time delay between the release of primary N from 
intermediate mass stars and O from short-lived massive stars.
The delay can be as large as $5\times 10^{8}$ yrs,
which is expected to introduce a large scatter in the N/O ratio 
at low O/H  (Garnett 1990;
Pilyugin 1993; Marconi, Matteucci, \& Tosi 1994). The scatter in N/O is also 
expected to increase
with decreasing metallicity as the effects of the time delay become 
relatively more important.

There are many observational investigations of the origin of N in the
literature. In particular, Vila-Costas and Edmunds (1993) 
summarized the then-existing observations of abundances in
H II regions in nearby spiral and dwarf irregular galaxies  
and found that at high metallicities
(O/H$\sim$ solar), the observed N/O ratio 
increases with O/H in a fashion consistent with the predictions 
from simple chemical evolution models. This indicates that secondary 
production of N dominates over primary production at high metallicities.
However, at low metallicities (O/H$<0.3$ solar or so), 
the N/O ratio becomes
roughly independent of O/H, indicating that primary N production dominates. 
There is also significant scatter ($\sim$ factor of 2) in the N/O ratio 
at low metallicities,  which cannot be attributed entirely to
measurement errors (Garnett 1990; Pagel et al 1992). The scatter can
instead be understood as the result of time delay between the injection of
primary N from intermediate mass stars and that of O from shorter-lived
massive stars  (cf. Vila-Costa \& Edmunds 1993).  
The trend of constant N/O (i.e., independent 
of O/H) in low metallicity H II regions has been confirmed
by more recent observations of blue compact and dwarf irregular galaxies,
some with even lower metallicities (Garnett 1990; Pagel et al 1992;
Skillman \& Kennicutt 1993; Thuan, Izotov, Lipovetsky 1995). 
However, the expected increase in the scatter of the N/O ratios 
toward decreasing metallicity is not observed (see figure 1).
In particular, Thuan et al (1995) called attention to the 
remarkably small scatter, 
0.08 dex rms, in the N/O ratios for the 15 blue compact galaxies in their 
sample with $1/30<$(O/H)$<1/6$ solar.  
Thuan et al argued that such a smaller scatter
in N/O is inconsistent with 
the standard time-delay model in which the release of primary N from 
intermediate mass stars is delayed relative to that of O from 
short-lived massive stars. Consequently, they argued that it must be 
possible to synthesize {\it primary} N in 
massive stars ($M>10M_{\odot}$). However, no clear mechanism for synthesizing
primary N in massive stars has been identified in the course of  
theoretical investigations (Woosley \& Weaver 1982; Maeder 1983).

The nearby blue compact and dwarf irregular galaxies exhibit a range in heavy 
element abundances because of their different chemical enrichment histories.
There is a rough correlation between the heavy element abundance and 
the luminosity of the system --- a fact which has been 
exploited in searches for dwarf, star-forming galaxies of very low 
metallicity.
However, for hitherto unexplained reasons, no nearby galaxy has 
been discovered with a lower heavy element abundance than that 
found in I Zw 18, namely, O/H$\sim$1/50 of the solar value.
 
A new approach to the problem of the origin of N was pioneered by 
Pettini, Lipman and Hunstead (1995) and by
Lipman, Pettini and Hunstead (1995), who examined the
behavior of the N abundance at large redshifts
in three damped Ly$\alpha$ (DLA) absorption systems in quasar spectra.
DLA systems are believed to be the high-redshift counterparts
of the present-day galaxies (see section 2 for more descriptions of their
basic properties and for references); some of them exhibit heavy element 
abundances as low as 1/200 solar at redshifts $z>2$ (cf. Lu et al 1996a). 
Consequently, they offer
the opportunity to examine the behavior of N in a hitherto unexplored
metallicity region and in stellar systems whose chemical enrichment
history may be simpler than that of nearby galaxies (note that the age
of the universe at $z=2$ is only $\sim$3 Gyrs). 
Unfortunately, the only accessible O I line, $\lambda$1302, is practically
always saturated even at the lowest abundance levels found in DLA systems.
Accordingly, it is only possible to derive lower limits to O/H
and upper limits to N/O for DLAs so that a direct comparison with the N/O
ratios measured in local metal-poor galaxies is not possible. 
For this reason, Pettini et al and Lipman et al used N/Si or N/S in 
place of N/O
on the ground that the O/Si and O/S ratios are found to be the same as
solar in Galactic disk and halo stars and
in H II regions of nearby galaxies.
While the small number of damped Ly$\alpha$
galaxies studied by these authors did not allow them to make any 
firm conclusions
regarding the origin of N, these studies did illustrate the potential
of the approach.

We describe measurements of N abundances in 15 damped Ly$\alpha$ galaxies
at $z>2$, largely based on high resolution, high S/N observations obtained
using the High Resolution Spectrometer (HIRES; see Vogt 1992) 
on the 10m Keck I telescope.
We demonstrate that the N/Si ratios in the most metal-poor 
([Si/H]$\sim 0.01$ solar) damped Ly$\alpha$ galaxies  do exhibit a very 
large scatter ($\sim 1$ dex) consistent with the prediction from
the time-delay model of primary N production. 
In Section 2 we describe the abundance measurements 
for N, O, Si, and S obtained for DLA systems. 
The analysis and discussion of the damped Ly$\alpha$ abundance
data in connection to the origin of N are presented in Section 3. We
give a brief summary of the results in Section 4.

\section{ABUNDANCE DATA FOR DAMPED LYMAN-ALPHA GALAXIES}

 We recall that DLA galaxies are the objects responsible
for producing neutral hydrogen absorption lines in the spectra of quasars 
with H I column densities $N$(H I)$>10^{20}$ cm$^{-2}$ (Wolfe 1988). 
The exact nature
of DLA galaxies remains unclear. They could be spirals,  dwarfs, or, at high 
redshifts, collapsing protogalaxies in the process of being assembled from 
smaller sub-units. Since the DLA clouds are very optically thick at Lyman 
limit, the fraction of ionized gas should be small.
DLA systems can currently be studied
over the redshift range $0.0 \le z \le 5$ and accurate abundances can be 
derived for many elements with minimal uncertainty in the ionization 
corrections. Accordingly, they provide the best opportunity
to directly probe the chemical evolutionary history of galaxies since very
early epochs.  General reviews of the properties of DLA absorption systems 
can be found in Wolfe (1988; 1993). Extensive discussions of elemental 
abundances in DLAs are provided by Pettini et al (1994; 1997a,b),
Lauroesch et al (1996), and by Lu et al (1996a).

The abundance of N has been measured for DLA galaxies only recently,
mostly based on HIRES echelle observations obtained with the
10m Keck I telescope.  A compilation of such measurements appears
in Table 1, where we give the quasar name and its emission redshift, 
the redshift and
H I column density of the DLA galaxy, 
and the abundances of O, N, Si, and S, as well as
references to the original sources of measurements. All
abundances are given relative to the solar values of Anders \& Grevesse
(1989) in the notation [M/H]=log(M/H)$_{DLA}-$log(M/H)$_{\odot}$. 
A few of the DLAs studied by Lu et al (1996a) did not have reported N
abundances because the N I lines are all contaminated by unrelated
Ly$\alpha$ forest absorption lines\footnote{The term ``Ly$\alpha$ forest''
refers to the region of a quasar 
spectrum blueward of the Ly$\alpha$ emission
line of the quasar, 
where hydrogen Ly$\alpha$ absorption lines arising from diffuse
intergalactic gas clouds at redshifts below that of the quasar litter the
quasar spectrum. These Ly$\alpha$ forest lines are relatively narrow 
(typical
Doppler width $20<b<50$ km s$^{-1}$) and weak, with 
neutral hydrogen 
column densities typically in the range $10^{12}<N$(H I)$<10^{15}$
cm$^{-2}$. The density of Ly$\alpha$ forest 
lines is very high, numbering in the hundreds 
per 1000 \AA\ at redshifts 2-4.}. However, it turns out that
useful limits on the N 
abundance can be obtained for these systems despite the
contaminations. We provide the N abundance estimates for these systems
in the Appendix and include the results in Table 1.  
Below we make some general remarks on the abundance measurements.

    The abundances of N  are usually determined from the N I
triplet absorption lines near $\lambda$1200 \AA\ and/or
the triplet near $\lambda$1134 \AA\ (especially
when the stronger triplet lines near $\lambda$1200 \AA\ are saturated).
Even though these  N I lines occur in the Ly$\alpha$ forest,
it is often still possible to reliably identify and measure the 
column density of these lines  because (1) there are three 
or more N I lines to work with; and (2) the N I lines are generally much
narrower than the Ly$\alpha$ forest lines owing to the larger mass
of N and perhaps to a lower temperature in the gas.
However, there are many cases where the N I lines are so weak and
the contamination from Ly$\alpha$ forest absorption is so strong
that no obvious N I absorption is discernible at any of the N I
line positions (e.g., the systems toward Q 1055+4611, Q 1202$-$0725, 
Q 2212$-$1626, Q 2233+1310, Q 2237$-$0608, and Q 2348$-$1444). 
In such cases, only upper limits to the N abundance are derived by
treating the contaminating Ly$\alpha$ forest absorption as N I
absorption (see examples in the Appendix). 
Although the statistical significances of such upper limits are
difficult to estimate, the true N abundances are likely to be
significantly lower than the upper limits provided in Table 1 for most cases.
There is also one case (the system toward Q 1425+6039) where the N I lines
are strongly saturated so that only a lower limit to the N abundance 
can be derived.

    The abundances of O are generally not measurable in DLA systems because
the accessible O I $\lambda$1302 absorption line is always saturated.
Consequently, only lower limits to O/H can be deduced. In some cases, 
the O I absorption line occurs in the Ly$\alpha$ forest 
and is severely affected by forest absorption
lines, making it impossible to derive even a lower limit to O/H.
In general, the O/H lower limits listed in Table 1 are not tight enough to
be of much use. The true O abundances are probably much higher than the 
lower limits listed in Table 1 in all cases given the heavy saturation
of the O I lines in these systems.

The abundances of Si are generally derived from unsaturated absorption lines
of Si II in the 
spectral region longward of the Ly$\alpha$ emission
lines. Hence they are not 
subjected to the contamination from Ly$\alpha$ forest lines.
When the Si II lines are saturated (7 cases in Table 1), 
only lower limits to the abundances of Si are provided
by assuming effectively the lines are not saturated.  
Component fitting analyses suggest that the true Si abundances
for 6 of the 7 lower limits given in Table 1 (Q1425+6039 being the
exception) are likely to be no more than a factor of 2-3 
(0.3-0.5 dex) higher than the lower limits provided.

The abundances of S are derived, where possible, 
from unsaturated absorption lines of the
S II triplet at $\lambda\lambda$1250, 1253, and 1259 \AA. These lines
usually occur in the Ly$\alpha$ forest and are often contaminated
by forest absorption lines. However, since there are three
lines to work with, it is often possible to reliably identify and
measure the column density of the lines. There are several cases where
the S II lines are either too weak or the contamination from the Ly$\alpha$
forest absorption is too strong that no obvious S II absorption is
discernible; only upper limits on the S abundance are provided in these
cases. Of the 4 upper limits listed in Table 1, the one for Q2344+1228
is a 2$\sigma$ upper limit, and the ones for Q1055+4611 and Q1946+7658
are based on measurements of the contaminating Ly$\alpha$ absorption
at the positions of the S II lines in a way similar to that described
in the Appendix.

\section{ANALYSIS AND DISCUSSION}

\subsection{Evidence for a Large Scatter in the N/Si Ratio}

The most useful way to discuss the origin of N production is to examine the
distribution of N/O as a function of O/H, as is normally done in 
such kinds of studies (eg, Vila-Costas \& Edmunds 1993). 
Constructing such a distribution for DLA systems is not literally possible
since only lower limits to O/H are available and,
in general, the O/H limits are not 
tight enough to provide significant constraints.
However, we note that the abundance of O has been found to trace that of
other $\alpha$-capture elements (e.g., Si, S) in both Galactic disk and halo
stars (cf. Wheeler, Sneden, \& Truran 1989) 
and in Galactic and extragalactic H II regions (cf. Thuan et al 1995 and
references therein) in the sense that [O/$\alpha$]$\simeq 0$
(i.e., O/$\alpha\simeq$solar).
Assuming the same holds true for DLA systems, 
we can substitute Si/H in place of O/H and substitute N/O with
N/Si. The results of this exercise are shown in figure 1, 
where the solid circles are [N/Si] vs [Si/H] for DLA systems
and the other data points are the actual [N/O] vs [O/H] measurements
from H II regions in nearby spiral, dwarf irregular, 
and blue compact galaxies 
(Garnett 1990; Pagel et al 1992; Vila-Costas \& Edmunds 1993;
Skillman \& Kennicutt 1993; Thuan et al 1995).
We also show the curves derived by Vila-Costas \& Edmunds (1993) 
indicating the contributions from primary 
and secondary N production.

     First, we note that the N/Si ratios in
DLAs occupy the same general region 
delineated by the ``secondary'' and ``primary+secondary''
curves that Vila-Costas \& Edmunds (1993) found to describe the N/O 
distribution
of H II region measurements. However, at the low metallicity end 
of the distribution ([Si/H]$\sim -2$), the N/Si ratio in DLAs
shows a considerably larger scatter ($\sim 1$ dex) than that exhibited
by nearby dwarf and blue compact galaxies of similar metallicities. 
We emphasize that the scatter is real. For example, of the 6 DLA
systems  at [Si/H]$\leq-1.8$, the highest {\it measured} value of [N/Si] 
is $-0.88$ while the lowest is $-1.70$. In addition, if we assume
[Si/H]$\simeq$[S/H], we find [N/Si]$\simeq-0.79$ for the Q1946+7658 
DLA system.
The large scatter in the N/Si ratios at very low metallicities
provides strong evidence that primary N production in DLA
galaxies does not go step-in-step with that of Si.
Rather, the large scatter is 
consistent with the long-sought behavior predicted
by the standard time-delay model of primary N production,
assuming O/Si$\simeq$ solar in DLA systems.
As we discuss in the next section, it is not possible to attribute
the large N/Si scatter in DLA systems to measurement uncertainties or 
possible systematic biases.

\subsection{Uncertainties}

There are a number of factors that could potentially affect the DLA 
abundance 
measurements and the interpretation of figure 1, including systematic 
biases due to line saturation effects, ionization effects, dust
depletion effects, and the assumption that O/Si$\simeq$solar in DLAs. 
We discuss each of them below, but note
that the latter two are likely to be the dominant sources of uncertainty.

(1) Most of the measurements given in Table 1 are based on high quality Keck
HIRES observations obtained by the present authors. We have tried to avoid
using saturated absorption 
lines to derive ion column densities. Other measurements quoted in
Table 1 were derived similarly.  Consequently, we do not
believe line saturation problems have significantly affected the abundance
measurements presented in Table 1.

(2) The abundances presented in Table 1 
were derived from the observed N I/H I, O I/H I, Si II/H I,
and S II/H I column density ratios under the
assumptions that the 
hydrogen gas in the DLA systems are mostly neutral and that
N I, O I, Si II, and S II are the dominant ionization stages 
of these elements in the gas.  
The rationale for making these assumptions is that the DLA systems
have Lyman limit optical depth $>10^3$; hence most of the elements should be
in the ionization stages that require more than 13.60 eV to ionize, namely,
N I (14.53 eV), O I (13.61 eV), Si II (16.34 eV), and S II (23.33 eV).
Simple photoionization calculations 
(Viegas 1995; Lu et al 1995, Prochaska \& Wolfe 1996)
support this conclusion.

(3) In order to compare the abundance measurements in DLAs with local
H II region measurements, it was assumed that O/Si$\simeq$solar in 
DLA systems.
The rationale for making this assumption was discussed in section 3.1.
It will be important to verify this
assumption observationally.

(4) The gas-phase abundance of heavy elements obtained from absorption 
line measurements may or may not reflect the total 
abundance of the elements in a galaxy, depending
on whether a significant fraction of the elements is locked up in dust 
grains\footnote{We ignore the heavy elements 
that may be incorporated in molecules
since few DLA systems appear to contain detectable H$_2$.}. 
There has been some debate about the extent to which the DLA abundances are
significantly affected by the effect of dust depletion 
(Pettini et al 1994,1997a; Lu et al 1996a; Lauroesch et al 1996;
Prochaska \& Wolfe 1996;
Kulkarni, Fall, \& Truran 1997; Welty et al 1997; Vladilo 1997).
The generally super-solar Zn/Fe and Zn/Cr ratios observed in DLA systems
have been interpreted to indicate a small amount of depletion of Fe and Cr
(cf. Pettini et al 1997a). On the other hand, the near-solar Si/S 
ratios found in DLAs (Table 1) seem to suggest that Si is not significantly
affected by dust depletion (Lu et al 1996a) because S is not readily
incorporated into dust grains in the Galactic ISM (Jenkins 1987).
We could have used S rather than Si in the analysis to avoid
the issue of dust depletion altogether; however, 
doing so would substantially reduce the sample size suitable for
this study. Interested readers are referred to these papers for detailed
discussions.  We merely note here that the effects of dust depletion
suggested to exist in DLA systems are not large enough to significantly 
alter our conclusions obtained in this study.  For example, 
the typical [Zn/Cr] and [Zn/Fe] values in DLAs at
$z>2$ are about $+0.4$ dex, 
with a range of 0 to $+0.65$ dex (Lu et al 1996a; Pettini et al 1997a). 
Zinc is largely unaffected by dust depletion effect in the Galactic ISM, 
while Cr and Fe are among the most refractory 
elements known (Jenkins 1987). Given that the relative abundances of
Zn, Cr, and Fe remain close to solar in Galactic stars with [Fe/H]$>-2.5$
(cf, Wheeler et al 1989), the DLA results suggest 
that Cr and Fe are depleted by about 0.4 dex (on average)
in these systems.  The depletion of Si in DLAs is expected to be 
less since Si is only moderately affected by dust in the
Galactic ISM (Jenkins 1987). 
For example, in ISM clouds where the depletion level of Fe and Cr is
similar to that inferred for DLA systems, Si is depleted by only
$\sim 0.26$ dex (see Table 7 of Sembach \& Savage 1996).
The abundances of N in DLAs should be largely
unaffected by dust depletion effect since in the Galactic ISM N 
is not readily 
incorporated into dust grains (cf. Savage \& Sembach 1996). 
{\it In summary, we expect dust depletion 
to shift the DLA measurements in figure 1 downward and to the right 
by an average of at most 0.3 dex.  Such a shift should
not affect the main conclusions of our analysis. 
In particular, the shift should 
not change appreciably the spread in N/Si}.

%

\subsection{Comparisons with Measurements in Nearby Galaxies}

     The result that DLAs show considerably 
larger scatter in their N/Si ratio than local galaxies with 
similar metallicities appears to have at least two possible explanations:

(1) It may be an unlucky consequence of small number statistics. 
Based on the limited DLA measurements available (see figure 1), 
it appears that the scatter in the N/Si ratio does not
become appreciably larger than that displayed by local H II region
measurements until [O/H] or [Si/H]$<-1.6$ or so. It may be significant
that only one local galaxies (I Zw 18) has metallicity
this low. Possibly, observations of more local galaxies with such
low metallicities will reveal a larger scatter in their N/O ratio.

(2) The DLAs may follow an intrinsically different evolution track
from the blue compact/dwarf irregular galaxies. In order to reproduce
simultaneously the observed abundances of He, N, and O in H II regions
of blue compact and dwarf irregular galaxies, 
several authors (eg, Pilyugin 1993;
Marconi et al 1994) found it necessary to include in the chemical
evolution calculations the effects of galactic winds powered by 
type II supernovae, which remove some of the SN II ejecta, rich in O and
$\alpha$ elements, from the galaxies but leave the abundance of N unaffected.
Variations in the efficiency of the galactic winds coupled with the
delayed release of N then create the scatter in the observed N/O ratios.
The very low N/Si ratios in some DLA galaxies may result if the effects
of galactic winds are less important for 
some reasons (eg, differences in galaxy masses or intensity of star
bursts).  Given the large number of parameters that go into chemical 
evolution
models (eg, the number, duration, and intensity of star bursts,
the mass of the galaxy,
the form of the stellar initial mass function, galactic winds, stellar
yields), it will be necessary to perform detailed calculations in order
to see if the above suggestion is tenable.

\section{SUMMARY}

 It is currently believed that primary N is produced by intermediate
mass (3-8$M_{\odot}$) stars during the Asymptotic Giant Branch phase, 
which dominates secondary N production at low metallicities.
Galactic chemical evolution model calculations indicate that, if the
above idea is correct, there should be considerable
scatter in the observed N/O ratio at a fixed metallicity (O/H) for galaxies
with low metallicities. The scatter stems from 
the delayed release of primary
N from intermediate mass stars relative to that of O from short-lived
massive stars. In addition, the scatter should increase progressively
toward decreasing metallicity. Such expected behaviors have not been
convincingly demonstrated by observations of H II regions in nearby
metal-poor galaxies. Consequently, several authors have suggested that
massive stars may dominate primary N production at low metallicities,
which is difficult to understand from a theoretical point of view.
We present observations of heavy element 
abundances in 15 high-redshift ($z>2$) damped Ly$\alpha$ galaxies, 
many of which have metallicities comparable to or lower than the
lowest-metallicity galaxy known locally (i.e., I Zw 18 with [O/H]$=-1.7$). 
We find that the N/Si ratio in our sample of damped Ly$\alpha$ galaxies
exhibit a very large scatter ($\sim1$ dex)
at [Si/H]$\sim -2$. Considerations of various sources of uncertainties
suggest that the large scatter is real.
These results provide strong support for
the time-delay model of primary N production in intermediate mass stars
if O/Si$\simeq$ solar in DLA galaxies. In particular, primary N production
in massive stars is not required. However, it will be important
to obtain accurate measures of O abundance in DLAs to verify the
assumption that O/Si$\simeq$solar in DLA galaxies.

\acknowledgements
The authors thank an anonymous referee for several insightful comments and
suggestions.
LL gratefully acknowledge support from NASA through grant number 
HF1062.01-94A from the
Space Telescope Science Institute, which is operated by the Association
of Universities for Research in Astronomy, Inc., for NASA under contract
NAS5-26555. WLWS acknowledges support from NSF grant AST95-29073.

\begin{planotable}{ccccrrrrrr}
\tablewidth{0pc}
\tablecaption{ABUNDANCES IN DAMPED LYMAN-ALPHA SYSTEMS}
\tablehead{
\colhead{QSO}              &\colhead{$z_{em}$}    
 &\colhead{$z_{DLA}$}      &\colhead{log N(HI)}
 &\colhead{[O/H]}          &\colhead{[N/H]} 
 &\colhead{[Si/H]}         &\colhead{[S/H]}      
 &\colhead{Ref}     }
\startdata
0000$-$2620 &4.108 &3.3901 &21.41 &\nodata$^a$ &$-2.77\pm0.17^a$  &$\geq-2.40$        &$\leq-1.98^e$   &1,2\nl
0100$+$1300 &2.702 &2.3090 &21.32 &$\geq-2.89$ &$-2.08\pm0.35$    &$-1.36\pm0.05$     &$-1.47\pm0.05$  &3\nl
0347$-$3819 &3.230 &3.0250 &20.70 &\nodata$^b$ &$-2.05\pm0.10^b$  &$\geq-1.43^b$      &\nodata         &9,10\nl
0930$+$2858 &3.433 &3.2353 &20.18 &$\geq-2.58$ &$-2.41\pm0.14$    &$\geq-1.79$        &$-1.78\pm0.15$  &3 \nl
1055$+$4611 &4.128 &3.3172 &20.34 &$\geq-2.16$ &$\leq-2.30$       &$\geq-1.60$        &$\leq-1.26$     &3\nl
1202$-$0725 &4.700 &4.3829 &20.60 &$\geq-2.03$ &$\leq-2.35$       &$-1.76\pm0.11$     &\nodata         &3,4\nl
1331$+$1704 &2.084 &1.7764 &21.18 &\nodata$^c$ &$-2.73\pm0.10^c$  &$-1.85\pm0.10$     &\nodata         &5,6,7 \nl
1425$+$6039 &3.173 &2.8268 &20.30 &\nodata     &$\geq-1.73$       &$\geq-1.07$        &\nodata         &1\nl
1946$+$7658 &2.994 &2.8443 &20.27 &$\geq-2.76^d$ &$-3.79\pm0.08^d$ &$-2.09\pm0.06^d$  &$\leq-1.79$     &3\nl
2212$-$1626 &3.992 &3.6617 &20.20 &$\geq-2.37$ &$\leq-2.67$       &$-1.90\pm0.08$     &\nodata         &1,3\nl
2233$+$1310 &3.299 &3.1493 &20.00 &$\geq-1.56$ &$\leq-1.73$       &$\geq-1.04$        &\nodata         &3 \nl
2237$-$0608 &4.559 &4.0803 &20.52 &\nodata     &$\leq-2.29$       &$-1.80\pm0.11$     &\nodata         &1,3\nl
2343$+$1230 &2.549 &2.4313 &20.34 &$\geq-1.93$ &$-1.72\pm0.08$    &$-0.74\pm0.11$     &$-1.90\pm0.08$  &3\nl
2344$+$1228 &2.773 &2.5379 &20.36 &$\geq-2.33$ &$-2.86\pm0.13$    &$\geq-1.67$        &$\leq-1.30$     &3\nl
2348$-$1444 &2.940 &2.2794 &20.57 &$\geq-2.71$ &$\leq-3.15$       &$-1.97\pm0.12$     &$-1.91\pm0.15$  &8\nl
\enddata

\end{planotable}

\noindent{\bf Notes to Table 1:}

\noindent$^a$ The [N/H] measurement is from Molaro et al (1996) 
based on three relatively weak N I 
lines near 1134 \AA. Molaro et al also quoted a [O/H]=$-3.13\pm0.17$ 
using an O I column density determined from fitting Voigt profiles to  a 
heavily saturated O I $\lambda$1302 absorption line. 
However, similar analysis of the same O I absorption line
by Lu et al (1996c) performed on superior quality Keck HIRES data
yielded a very different O I column density. We believe saturated absorption
lines can not be used to derive accurate column densities and abundances.

\noindent$^b$ The [N/H] measurement is 
from Vladilo et al (1997),
who also quoted a [O/H]$=-1.12\pm0.31$ for this system.  However,
the latter measurement was based on profile fitting to a saturated O I 
absorption line; we prefer the more conservative limit of [Si/H]$>-1.43$
from Wolfe \& Prochaska (1997, private communication) in the analysis.

\noindent$^c$ The [N/H] is derived using a 
log $N$(H I)=21.18 from Wolfe (1995) and a log $N$(N I)=$14.5\pm0.1$ 
derived by Green et al (1995) from three mildly saturated N I lines
near 1200 \AA. Green et al also quoted a log $N$(O I)$\sim$15.3, 
which implies a [O/H]$\sim-2.81$ for this system. 
However, the O I column density was derived from fitting
Voigt profiles to a strongly saturated O I $\lambda$1302 
absorption line and is therefore extremely uncertain.

\noindent$^d$ Similar measurements were made by Lu et al (1996a). We
update the measurements here based on higher S/N Keck HIRES observations.

\noindent$^e$ An abundance of [S/H]$=-1.98$ is obtained (ref 1) if
the absorption feature tentatively identified as S II $\lambda$1253
by Lu et al (1996c) is indeed S II. However, the absorption feature
could also be a contaminating Ly$\alpha$ absorption line. Hence we
conservatively list the value as an upper limit.

\bigskip\bigskip
\noindent{\bf References to Table 1: }
(1) Lu et al 1996a;
(2) Molaro et al 1996;
(3) This paper; Details of the measurements will be provided 
    elsewhere (Lu et al 1998, in preparation).
(4) Lu et al 1996b;
(5) Green et al 1995;
(6) Wolfe 1995;
(7) Kulkarni et al 1996;
(8) Pettini, Lipman, \& Hunstead 1995;
(9) Vladilo, Matteucci, \& Molaro 1996;
(10) Wolfe \& Prochaska 1997 (private communication).

\clearpage
\begin{figure}
\plotone{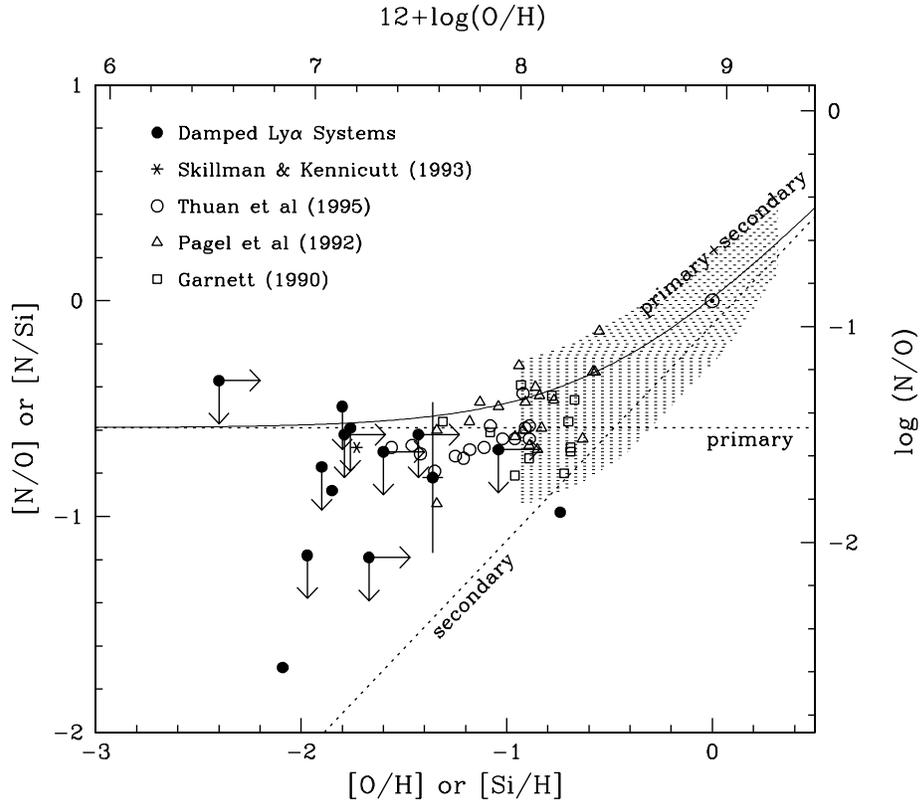}
\figcaption{This figure shows [N/Si] vs [Si/H] for damped Ly$\alpha$ 
galaxies (solid circles)  and [N/O] vs [O/H] for nearby dwarf irregular 
and blue compact galaxies (other symbols).  
Measurement uncertainties (not shown) are typically
$\leq 0.1$ dex for the damped Ly$\alpha$ systems unless otherwise indicated.
The shade area indicates the region occupied by 
the [N/O] vs [O/H] measurements in H II regions of nearby spirals 
based on the 
compilation of Vila-Costas \& Edmunds (1993).  The measurement 
in the Sun is indicated by the Sun ($\odot$) symbol. The two dashed
lines indicate the contributions from primary and secondary N 
productions determined by Vila-Costas \& Edmunds (1993), and the solid curve
is the sum of two.  The top horizontal axis and the right vertical axis 
show  the usual scale of 12+log(O/H) vs log(N/O).}
\end{figure}

\clearpage
\centerline{\bf APPENDIX}

    Three of the DLA systems listed in Table 1, 
the $z=4.3829$ system toward Q 1202$-$0725, the $z=3.6617$ system 
toward Q 2212$-$1626, and the $z=4.0803$ system toward Q 2237$-$0608,
were included in the 
analysis of Lu et al (1996a,b), who did not provide measurements of the
N abundance in these systems 
because the spectral regions where the N I lines occur were
badly contaminated  by Ly$\alpha$ 
forest absorption lines. However, it turns out that useful
upper limits on the N abundance for these systems can be derived despite the
contaminations. We describe these estimates below. 
All data presented here were based on Keck I HIRES observations 
at 6.6 km s$^{-1}$ resolution  and have been described in details 
elsewhere (see Lu et al 1996a,b).

    Figure 2 shows the spectral regions near the N I 
$\lambda\lambda$1199.55, 1200.22, and 1200.71 
lines in the $z=4.3829$ DLA system toward Q 1202$-$0725,
plotted in velocity space in the rest-frame of the DLA galaxy.
Also shown is the 
well-observed absorption profile of the Si II $\lambda$1304 line
in the same system  to help define the region where N I absorption 
may be expected to occur. The contamination from the Ly$\alpha$
forest absorption lines is strong 
enough and the N I lines appear weak enough
that no discernible evidence of N I absorption is present. Direct
integrations of the apparent 
column density profiles ($N_a(v)$; see Lu et al 1996a for detailed
description) 
of the three N I lines over the velocity interval [$-30$,160] km s$^{-1}$,
where most of the N I absorption is expected to occur 
based on the Si II $\lambda$1304 absorption, yield the following 
column densities: $>14.96$ ($\lambda$1199.55), 14.22 ($\lambda$1200.22),
and 14.96 ($\lambda$1200.71). 
The value for $\lambda$1199.55 is a lower limit because the
absorption is saturated.
These values should be considered strictly as upper limits to
the actual N I column density since 
most or all of the absorption included in the
integration regions is due to Ly$\alpha$ forest absorption. 
Taking the upper limit from the $\lambda$1200.22 line, 
we find log $N$(N I)$<14.22$ and [N/H]$<-2.35$ adopting the
$N$(H I) from Lu et al (1996b).

    Figure 3 shows the similar plots for 
the $z=3.6617$ DLA system toward Q 2212$-$1626.  Again, the regions near 
the N I triplet lines are contaminated by Ly$\alpha$ forest absorption
lines and there is no obvious evidence that N I absorption is present. 
The best upper limit on $N$(N I) is provided by the 
intrinsically strongest $\lambda$1199.55 absorption line,
which yields log $N$(N I)$<13.64$ 
over the velocity range [$-$50, +50] km s$^{-1}$.
We thus find [N/H]$<-2.67$ for this system adopting the $N$(H I)
from Lu et al (1996a).

   Figure 4 shows the similar plots 
for the $z=4.0803$ DLA system toward Q 2237$-$0608.
Again, no evidence for the presence of N I absorption is evident. 
In this case, the best upper limit on $N$(N I) is obtained by 
combining the $N_a(v)$ integration from N I $\lambda$1200.22
over [$-$90,+60] km s$^{-1}$ with 
that from the $\lambda$1200.71 line over [$-$130,$-$90]
km s$^{-1}$. We find log $N$(N I)$<14.28$ and [N/H]$<-2.29$
adopting the $N$(H I) from Lu et al (1996a).

\clearpage
\begin{figure}
\plotone{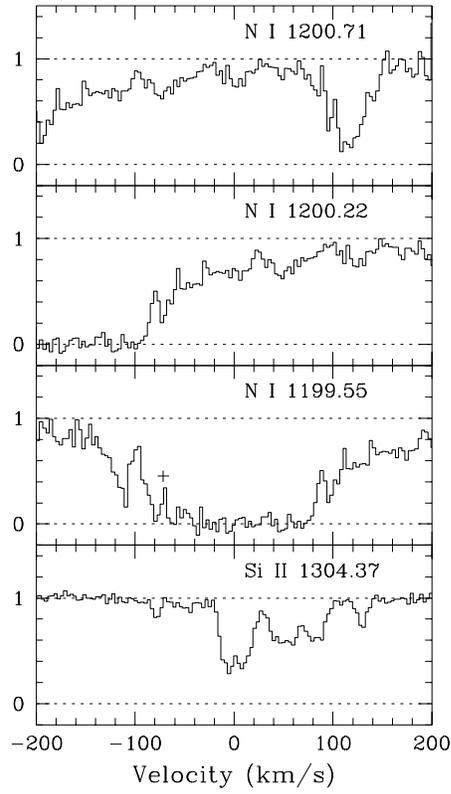}
\figcaption{Spectral regions near N I absorption lines in 
the $z=4.3829$ DLA system toward Q 1202$-$0725, 
plotted in velocity space in the rest-frame of the DLA galaxy.
These spectral regions are 
contaminated by Ly$\alpha$ forest absorption lines.
The absorption profile of Si II $\lambda$1304 in the same system is 
shown for comparison.}
\end{figure}

\clearpage
\begin{figure}
\plotone{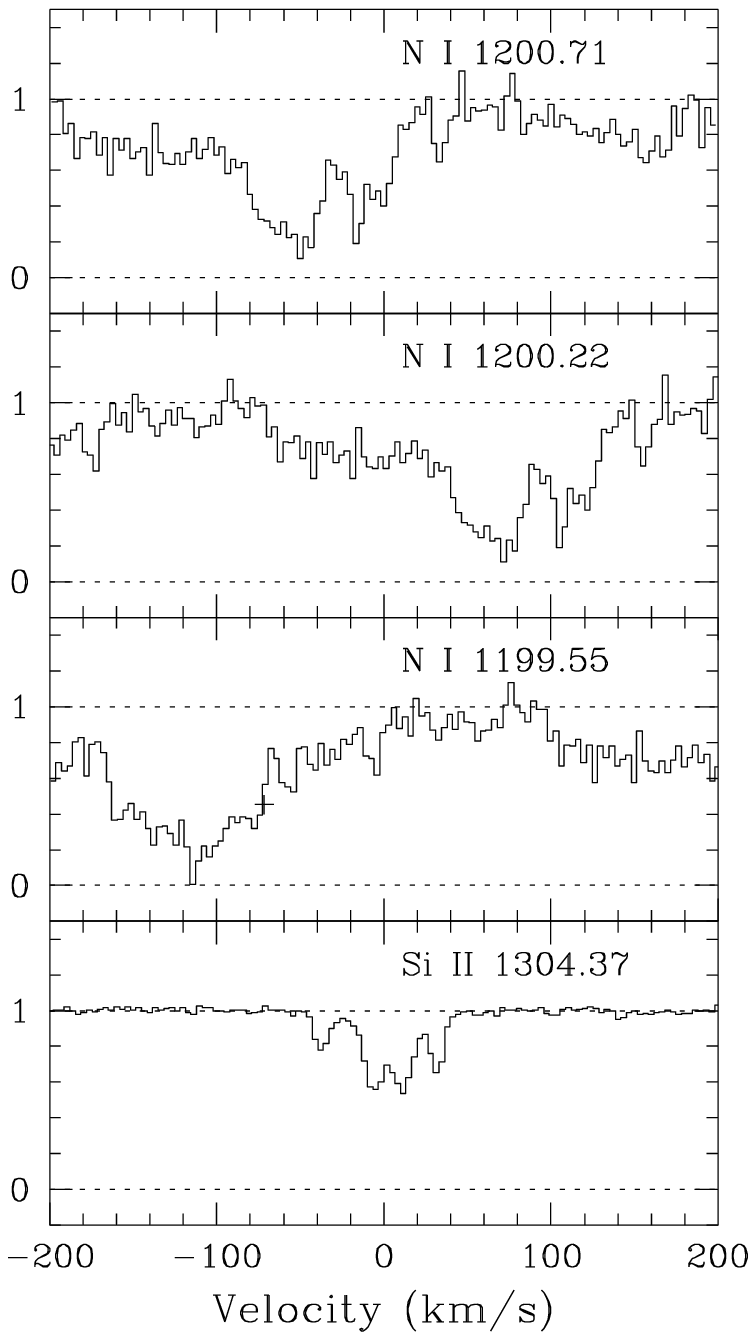}
\figcaption{Spectral regions near N I 
absorption lines in the $z=3.6617$ DLA system toward
Q 2212$-$1626, plotted in velocity 
space in the rest-frame of the DLA galaxy.  These spectral regions 
are contaminated by Ly$\alpha$ forest absorption lines.
The absorption profile of Si II $\lambda$1304 in the same system is 
shown for comparison.}
\end{figure}

\clearpage
\begin{figure}
\plotone{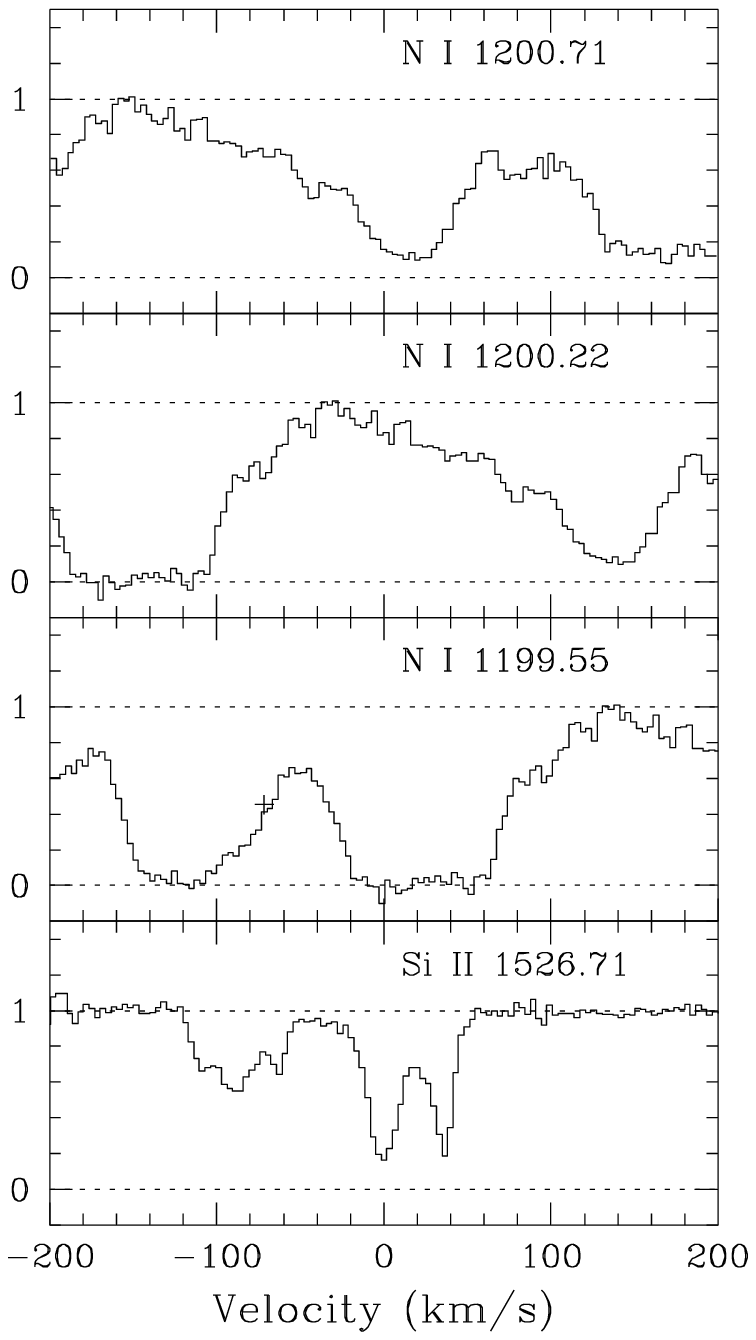}
\figcaption{Spectral regions near N I absorption 
lines in the $z=4.0803$ DLA system toward Q 2237$-$0608, 
plotted in velocity space in the rest-frame of the DLA galaxy.
These spectral regions are 
contaminated by Ly$\alpha$ forest absorption lines.
The absorption profile of Si II $\lambda$1526 in the same system is 
shown for comparison.}
\end{figure}

\end{document}